\documentclass[useAMS,usenatbib,usegraphicx]{mn2e}
\def\figscl{0.9}

\usepackage{color}

\usepackage[colorlinks,citecolor=black,linkcolor=black,urlcolor=black]{hyperref}
\newcommand*{\figref}[1]{Figure~\ref{#1}}
\newcommand*{\figrefs}[1]{Figures~\ref{#1}}
\newcommand*{\tabref}[1]{Table~\ref{#1}}
\newcommand*{\tabrefs}[1]{Tables~\ref{#1}}

\newcommand*{\secref}[1]{Section~\ref{#1}}
\newcommand*{\secrefs}[1]{Sections~\ref{#1}}

\newcommand*{\unit}[1]{\ensuremath{\mathrm{\, #1}}}

\newcommand*{\eV}{\unit{eV}}
\newcommand*{\second}{\unit{s}}

\newcommand*{\km}{\unit{km}}
\newcommand*{\Mpc}{\unit{Mpc}}

\newcommand*{\mysub}[2]{\ensuremath{#1_{\mathrm{#2}}}}

\newcommand*{\Omegam}{\mysub{\Omega}{m}}
\newcommand*{\Omegab}{\mysub{\Omega}{b}}

\newcommand*{\Omegak}{\mysub{\Omega}{k}}
\newcommand*{\LCDM}{\ensuremath{\Lambda}CDM}

\newcommand*{\fgas}{\mysub{f}{gas}}
\newcommand*{\Mgas}{\mysub{M}{gas}}
\newcommand*{\Mtot}{\mysub{M}{tot}}

\newcommand*{\Neff}{\mysub{N}{eff}}
\newcommand*{\Mnu}{\ensuremath{M_\nu}} 
\newcommand*{\nt}{\mysub{n}{t}}
\newcommand*{\ns}{\mysub{n}{s}}
\newcommand*{\zeq}{\mysub{z}{eq}}

\defcitealias{Mantz09}{Paper~I}
\newcommand*{\cosmopaper}{\citetalias{Mantz09}}
\defcitealias{Mantz09a}{Paper~II}
\newcommand*{\scalingpaper}{\citetalias{Mantz09a}}

\newcommand*{\Chandra}{{\it{Chandra}}}
\newcommand*{\ROSAT}{{\it{ROSAT}}}
\newcommand*{\WMAP}{{\it{WMAP}}}

\title[Cluster growth: neutrino properties]{The Observed Growth of Massive Galaxy Clusters IV: Robust Constraints on Neutrino Properties}
\author[A. Mantz et al.]{
  A.~Mantz,$^{1,2,3}$\thanks{E-mail: {\tt adam.b.mantz@nasa.gov}} S.~W.~Allen$^{2,3}$ and D.~Rapetti$^{2,3}$\\
  $^1$NASA Goddard Space Flight Center, Code 662, Greenbelt, MD 20771, USA\\
  $^2$Kavli Institute for Particle Astrophysics and Cosmology, Stanford University, 452 Lomita Mall, Stanford, CA 94305-4085, USA\\
  $^3$SLAC National Accelerator Laboratory, 2575 Sand Hill Road, Menlo Park, CA 94025, USA
}
\date{Accepted 2010 April 6. Received 2010 March 14; in original form 2009 November 5}

\begin{document}
\pagerange{\pageref{firstpage}--\pageref{lastpage}} \pubyear{2010}
\maketitle
\label{firstpage}

\begin{abstract}
  This is the fourth of a series of papers in which we derive simultaneous constraints on cosmological parameters and X-ray scaling relations using observations of the growth of massive, X-ray flux-selected galaxy clusters. Our data set consists of 238 clusters drawn from the \ROSAT{} All-Sky Survey, and incorporates extensive follow-up observations using the \Chandra{} X-ray Observatory. Here we examine the constraints on neutrino properties that are enabled by the precise and robust constraint on the amplitude of the matter power spectrum at low redshift available from our data. In combination with cluster gas mass fraction, cosmic microwave background, supernova and baryon acoustic oscillation data, and incorporating conservative allowances for systematic uncertainties, we limit the species-summed neutrino mass, \Mnu{}, to $<0.33\eV$ at 95.4 per cent confidence in a spatially flat, cosmological constant (\LCDM) model. In a flat \LCDM{} model where the effective number of neutrino species, \Neff{}, is allowed to vary, we find $\Neff=3.4_{-0.5}^{+0.6}$ (68.3 per cent confidence, incorporating a direct constraint on the Hubble parameter from Cepheid and supernova data). We also obtain results with additional degrees of freedom in the cosmological model, in the form of global spatial curvature (\Omegak) and a primordial spectrum of tensor perturbations ($r$ and \nt{}). The results are not immune to these generalizations; however, in the most general case we consider, in which \Mnu{}, \Neff{}, curvature and tensors are all free, we still obtain $\Mnu<0.70\eV$ and $\Neff=3.7 \pm 0.7$ (at, respectively, the same confidence levels as above). These results agree well with recent work using independent data, and highlight the importance of measuring cosmic structure and expansion at low as well as high ($z \sim 1100$) redshifts. Although our cluster data extend to redshift $z=0.5$, the direct effect of neutrino mass on the growth of structure at late times is not yet detected at a significant level.
\end{abstract}

\begin{keywords}
  cosmology: observations -- cosmological parameters -- large-scale structure of Universe -- X-rays: galaxies: clusters.
\end{keywords}

\section{Introduction} \label{sec:introduction}

Observations of neutrino flavor oscillation have conclusively shown that the neutrino mass eigenstates are non-degenerate \citep[e.g.][]{Fukuda98,Fukuda02,Ahmad02,Ahn03,Ahn06,Eguchi03,Sanchez03,Giacomelli04,Aharmim05}. Although, these observations can place tight constraints on the differences in squared mass, measuring the absolute mass scale remains challenging. Current laboratory efforts focus on tritium beta decay \citep[e.g.][]{Lobashev03,Kraus05} and neutrinoless double beta decay \citep[e.g.][]{Aalseth99,Klapdor01,Arnaboldi05,Arnold05}; the latter, if observed, would additionally indicate that neutrinos are Majorana rather than Dirac fermions. The squared mass differences measured from flavor oscillations place a lower bound on the sum of the three masses, $\Mnu=\sum_i m_i$, at $\sim 0.056$ (0.095)$\eV/c^2$ in the normal (inverted) hierarchy, while current tritium beta decay results provide an upper bound on the mass of the electron neutrino at $\sim 2\eV/c^2$ (thus on \Mnu{} at $\sim 6\eV/c^2$).\footnote{Henceforth, we set $c=1$.}

Because neutrinos play a prominent role in the early Universe, cosmological observations are also sensitive to their properties (for a review, see \citealt{Lesgourgues06}; see also \secref{sec:background}). The primary effect of non-zero neutrino mass on cosmological observables is to suppress the formation of cosmic structure on intermediate and small scales. Comparison of the Cosmic Microwave Background (CMB), which reflects large-scale structure at early times, with measurements of the intermediate- or small-scale structure in the local Universe thus provides a way to constrain the absolute mass scale of the neutrino \citep[e.g.][]{Fukugita00}.

This approach has the disadvantage that any neutrino properties inferred are at some level sensitive to our incomplete understanding of cosmology, in particular dark energy and inflation. It has long been recognized, however, that using complementary measurements of structure, as described above, significantly reduces the sensitivity of the results to such necessary assumptions (e.g. \citealt*{Allen03a}; \citealt{Tegmark04}). Nevertheless, it is imperative that any systematic uncertainties affecting the measurements of cosmic structure be properly accounted for in order to obtain robust results.

The amount of structure in the local Universe is generally described by the parameter $\sigma_8$, defined as the present day root-mean-square fluctuation of the linearly evolved matter density field, smoothed by a spherical top-hat window of comoving radius $8h^{-1}\Mpc$; here $h=H_0/100\km\second^{-1}\Mpc^{-1}$ is the normalized Hubble parameter. At present, the most robust observation for measuring this quantity is arguably the abundance of massive galaxy clusters;\footnote{We note that current cluster measurements do not constrain $\sigma_8$ independent of spectral index of the power spectrum, \ns{}. However, the best fitting value of $\sigma_8$ does not vary rapidly with \ns{}; moreover, \ns{} is well constrained by CMB data in all cosmological models considered here.} recent advances in cluster simulation \citep*[e.g.][]{Nagai07}, comparisons of different mass measurement techniques \citep[e.g.][]{Bradac08,Mahdavi08,Newman09}, and the availability of robust mass proxies (e.g. \citealt{Allen04,Allen08}; \citealt*{Kravtsov06}) have significantly reduced systematic uncertainties associated with the measurement of cluster masses. Consequently, recent estimates of $\sigma_8$ based on independent analyses of galaxy cluster data (both optically and X-ray selected clusters) are in very good agreement [$\sigma_8 \sim 0.8$; \citealt{Mantz08,Mantz09} (hereafter \cosmopaper{}); \citealt{Henry09,Vikhlinin09a,Rozo10}]. Recent analyses of cosmic shear data are also in good agreement \citep{Benjamin07,Fu08}, and provide compatible constraints on the neutrino mass to our own \citep{Tereno09}. Galaxy redshift surveys have also produced comparable results \citep*{Thomas09}.

The number of neutrino species participating in weak interactions is known to be three to high precision \citep{Amsler08}. However, the possibility remains that additional ``sterile'' species exist. Results consistent with the existence of a fourth neutrino were reported by the Liquid Scintillator Neutrino Detector \citep[LSND;][]{Aguilar01}; these results are disfavored by the MiniBooNE experiment \citep{MiniBooNE09,MiniBooNE09a}, although the interpretation remains somewhat ambiguous at present. This is another question that cosmological observations can address. In particular, the synthesis of light elements is sensitive to the number of relativistic species present in the early Universe, since these determine the expansion rate at that time; however, observations of primordial deuterium and helium abundances are challenging and are subject to large systematic uncertainties. An independent probe is provided by the CMB, in combination with other cosmological data, as described in \secref{sec:background}. In this work, we consider only the case where the neutrino species (whatever their number) have approximately degenerate mass; in this case, \Mnu{} and the effective number of neutrino species, \Neff{}, are sufficient to describe the cosmological effect of neutrinos. More general mass splittings, and in particular the case favored by the initial LSND results, in which a sterile neutrino is significantly more massive than the other species, require a more complete treatment \citep[as in, e.g.,][]{Crotty04}.

In this paper, we apply the statistically rigorous analysis method of \cosmopaper{} and the X-ray flux-limited cluster samples and follow-up observations described in \citet[][hereafter \scalingpaper{}]{Mantz09a} to the problem of inferring neutrino properties from cosmological data. These data (collectively referred to here as the cluster X-ray Luminosity Function, or XLF) provide a robust means to measure $\sigma_8$; our analysis method includes generous systematic allowances and accounts fully for all parameter degeneracies. In obtaining our results, we also incorporate CMB data and measurements of cosmic distance in the form of cluster gas mass fractions (\fgas{}), type Ia supernova (SNIa) fluxes and Baryon Acoustic Oscillation (BAO) data (see \secref{sec:data} and references therein). We note that \citet{Reid09a} obtain very similar results to our own by importance sampling 5-year {\it Wilkinson Microwave Anisotropy Probe} (\WMAP{}) results using a prior based on the analysis of optically selected clusters by \citet{Rozo10}.

The basic cosmology that we consider in this paper is the spatially flat, cosmological constant (\LCDM{}) model parametrized by the mean baryon density, \Omegab{}; the mean total matter density including baryons, neutrinos and cold dark matter (CDM), \Omegam{}; the Hubble parameter, $H_0$; the normalization of the matter power spectrum, $\sigma_8$; the spectral index of the primordial scalar power spectrum, \ns{}; and the optical depth to reionization, $\tau$. Here the mean densities refer to the present day (redshift $z=0$), since their values at other times are then determined by the Friedmann equation, and $\sigma_8^2$ is the $z=0$ variance in the matter density field at scales of $8h^{-1}\Mpc$, as defined above. This simple and commonly used model assumes $\Mnu=0$ and $\Neff=3.046$, the predicted value for the three weakly interacting neutrinos.

In addition to freeing \Mnu{} and \Neff{}, we will consider the effect of marginalizing over other parameters with which they are degenerate. However, since the flat \LCDM{} model is known to provide a good fit to currently available cosmological data, we are conservative in incorporating these additional degrees of freedom. In particular, we consider generalizing the description of dark energy, either by allowing global spatial curvature or by varying the dark energy equation of state in a flat universe, and including the effects of primordial tensor perturbations. The former case is parameterized by the effective curvature energy density, \Omegak{}, or the equation of state parameter, $w$, while in the latter case we simultaneously marginalize over the tensor-to-scalar ratio, $r$ (defined with respect to wavenumber $k=0.002h\Mpc^{-1}$, as in, e.g.,  \citealt{Komatsu09}), and the tensor power spectral index, \nt{}.

In our results, we will consistently quote one-sided limits (upper bounds) on \Mnu{} and $r$ at the 95.4 per cent confidence level and two-sided constraints on all other parameters at 68.3 per cent confidence.

\section{Background} \label{sec:background}

Although the purpose of this work is to explore the constraints enabled by the XLF data, it is instructive to begin by considering the cosmological constraints on \Mnu{} that have been obtained without using measurements of local cosmic structure. As described by \citet*{Ichikawa05}, \citet{Dunkley09} and \citet{Komatsu09}, CMB observations alone provide a limit that, while weak ($\Mnu<1.3\eV$ at 95.4 per cent confidence), is relatively robust to assumptions about dark energy and primordial tensors.\footnote{We note that this constraint is not entirely robust, and can be significantly degraded when multiple additional degrees of freedom are included in the cosmological model. In several of the models we consider, the combination of CMB data and cosmic distance measurements produces weaker constraints; however, in the worst case, the upper limit is still $\sim 2.5\eV$ (\tabrefs{tab:msdegen} and \ref{tab:nhdegen}).} While we do not review the details here, the constraint arises because the effect of neutrino mass on the CMB temperature anisotropy spectrum is qualitatively different when neutrinos are relativistic compared with non-relativistic during the decoupling epoch. In particular, for $\Mnu<1.5\eV$, changes in \Mnu{} can be easily mimicked by a corresponding change in the Hubble parameter, $H_0$. As a result, \Mnu{} is strongly degenerate with $H_0$ in analyses using only CMB data, which does not itself constrain $H_0$. The combination of CMB data with distance measurements in the form of SNIa and BAO data can place a constraint on $H_0$, improving the limits on \Mnu{} by roughly a factor of two \citep[$\Mnu<0.67\eV$ for a spatially flat \LCDM{} background;][]{Komatsu09}.

As we will show, the inclusion of a measurement of local cosmic structure in the form of galaxy cluster abundance can improve these limits by another factor of two or greater, and furthermore greatly increase the robustness of the results to assumptions about the nature of dark energy and inflation. This improvement is possible because the combination of large-scale, high-redshift and intermediate-scale, low-redshift measurements of cosmic structure exploits the defining characteristic of light, weakly interacting particles: they are relativistic during the earliest stages of cosmic structure formation but non-relativistic at the present day.\footnote{The latter condition follows from the lower bound $\Mnu>0.056\eV$ obtained from flavor oscillation measurements.} The imprint of neutrinos on cosmic structure, and in particular on the growth of structure from the surface of last scattering to the present, thus differs from that of the photon background (relativistic at all times) and CDM (assumed to be non-relativistic at all times). In particular, a non-zero mass in neutrinos results in a net suppression of the growth of structure on scales smaller than their free-streaming length (approximately corresponding to wavenumber $10^{-2}h\Mpc^{-1}$ today) relative to an equal mass in CDM \citep{Bond80}.

\begin{figure*}
  \centering
  \includegraphics[scale=\figscl]{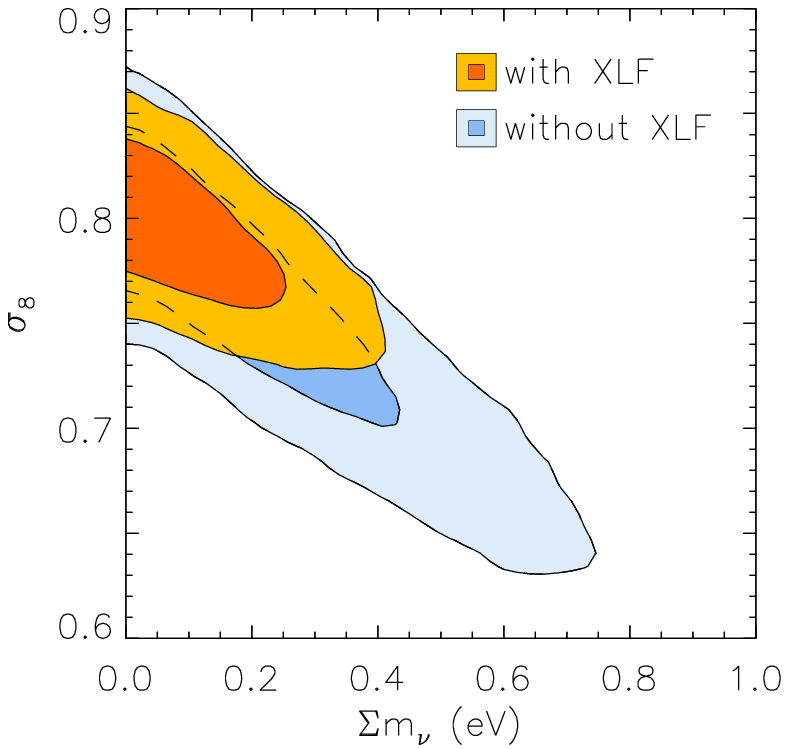}
  \hspace{0.5cm}
  \includegraphics[scale=\figscl]{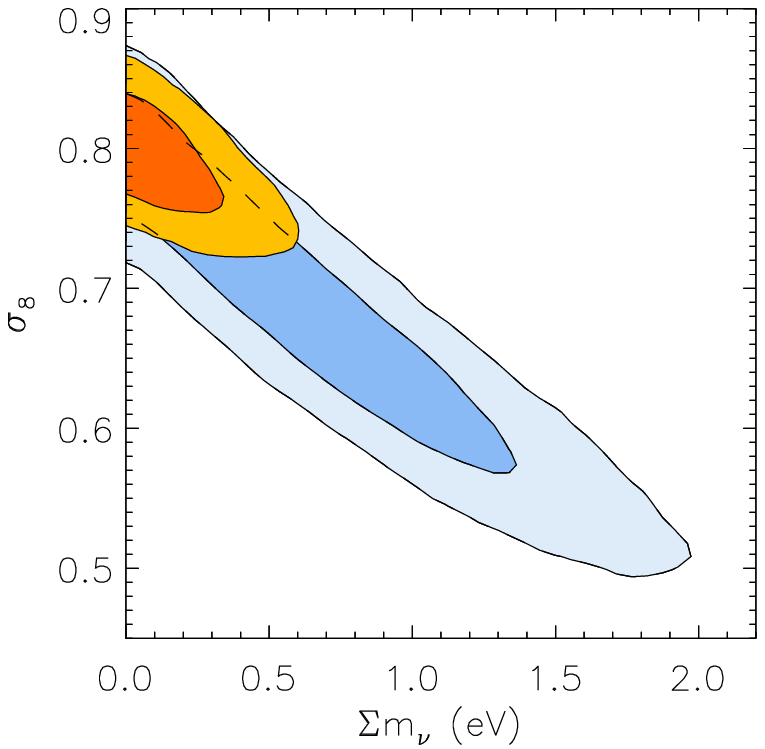}
  \caption{Joint 68.3 and 95.4 per cent confidence regions in the \Mnu-$\sigma_8$ plane from the combination of CMB, \fgas{}, SNIa and BAO data (blue), and the combination of those data with the XLF (gold). The XLF data provide a tight constraint on $\sigma_8$, breaking the degeneracy in this plane. To demonstrate the robustness of constraints on \Mnu{} obtained using the XLF data, we compare (left panel) results for a simple \LCDM+\Mnu{} model with (right panel) results obtained when nuisance parameters are included in the model (in this case, \Omegak{}, $r$ and \nt{}; note the difference in scale). Note that conservative systematic uncertainties are included here (and in all subsequent figures).}
  \label{fig:mslcdm}
\end{figure*}

In practical terms, the constraint on \Mnu{} arising from the combination of CMB and cluster data can be understood as follows. Even when $\Mnu \neq 0$, the CMB data provide a good constraint on the high-redshift amplitude of the power spectrum, since the direct effect of neutrino mass on the observed temperature anisotropies is relatively small. However, the late-time, intermediate-scale power spectrum amplitude (the value of $\sigma_8$) that would be predicted based on that high-redshift constraint is very sensitive to \Mnu{}, since neutrino mass suppresses the development of structure on those scales. This strong degeneracy is apparent in the blue confidence regions in \figref{fig:mslcdm}. By constraining $\sigma_8$ independently of \Mnu{}, clusters (or other low-redshift structure measurements) break this degeneracy, improving the constraints on neutrino mass (gold contours in the figure).

Even more power is available if the growth of structure can be measured, exploiting the time-dependent nature of the effects of neutrino mass on cosmic structure development \citep[e.g.][]{Lesgourgues06}. However, we show in \secref{sec:mnusimple} that the growth of structure in current cluster data, while able to constrain models of dark energy \citep[\cosmopaper{};][]{Vikhlinin09a} and modified gravity (\citealt{Rapetti09,Rapetti09a}; \citealt*{Schmidt09}), are not yet sufficient to contribute to constraints on \Mnu{}.

In the case of the effective number of relativistic species, \Neff{}, CMB data alone can provide a lower bound, due to the gravitational effect of their anisotropic stress. Apart from providing this bound, however, the CMB data are hampered by a near-perfect degeneracy between \Neff{} and $\Omegam h^2$, the physical matter density \citep{Dunkley09}. This degeneracy arises because CMB observations do not constrain either \Neff{} or $\Omegam h^2$ directly, only the redshift of matter-radiation equality, \zeq{}, which is a function of both parameters. Again, the addition of cosmic distance measurements can reduce the effect of this degeneracy by placing an independent constraint on $\Omegam h^2$; \citet{Komatsu09} obtained $\Neff=4.4 \pm 1.5$ (68 per cent confidence).

Measurements of local structure can also contribute to improved constraints on \Neff{}, as we show in \secref{sec:neffresults}. The contribution of clusters to this problem is less direct than in the case of \Mnu{}: the constraint on $\sigma_8$ and \Omegam{} from cluster data improves the constraint on $\Omegam h^2$ from the CMB+distance measurements (e.g. Figure~1 of \cosmopaper{}), indirectly improving the constraint on \Neff{} through the degeneracy described above.

\section{Data} \label{sec:data}

The galaxy cluster data used in this work, as well as their selection and reduction, are discussed in detail in \scalingpaper{}. Using three wide-area cluster samples drawn from the \ROSAT{} All-Sky Survey \citep[RASS;][]{Trumper93} -- the \ROSAT{} Brightest Cluster Sample \citep[BCS;][]{Ebeling98}, the \ROSAT{}-ESO Flux-Limited X-ray sample \citep[REFLEX;][]{Bohringer04}, and the bright sub-sample of the Massive Cluster Survey  (Bright MACS; \citealt*{Ebeling01}; \citealt{Ebeling10}) -- we select a statistically complete sample of 238 X-ray luminous clusters covering the redshift range $z<0.5$. Of these 238 clusters, 94 have follow-up \Chandra{} or \ROSAT{} observations that we incorporate into the analysis. From the follow-up observations, we measure X-ray luminosity, average temperature, and gas mass within $r_{500}$.\footnote{$r_{500}$ is defined to be the radius within which the mean enclosed density is 500 times the critical density at the cluster's redshift.} The gas mass is used as a proxy for total mass, using the finding of \citet{Allen08} that the gas mass fraction, $\fgas=\Mgas/\Mtot$, is a constant for hot, massive clusters (see also \secref{sec:method}). Cluster \fgas{} measurements additionally provide a precise measure of cosmic distance, so we include the full data set and analysis of Allen et~al. in this work.

In addition to these cluster data sets, we incorporate CMB, SNIa and BAO data. Our analysis of the CMB anisotropies uses 5-year \WMAP{} data \citep{Hinshaw09,Hill09,Nolta09} with the March 2008 version of the \WMAP{} likelihood code\footnote{\url{http://lambda.gsfc.nasa.gov}} \citep{Dunkley09}, as well as Arcminute Cosmology Bolometer Array Receiver (ACBAR) data at smaller angular scales \citep{Reichardt09} and measurements of the CMB polarization from the Background Imaging of Cosmic Extragalactic Polarization (BICEP) instrument \citep{Chiang09}. The SNIa results are derived from the Union compilation \citep{Kowalski08}, which includes data from a variety of sources (307 SNIa in total; see references in \cosmopaper{}). Our analysis of BAO data uses the constraints on the ratio of the sound horizon to the distance scale at $z=0.25$ and $z=0.35$ derived by \citet{Percival07} from the galaxy correlation function in 2dF \citep{Colless01,Colless03} and Sloan Digital Sky Survey \citep{Adelman-McCarthy07} data. Finally, we employ in some cases a Gaussian prior on the Hubble parameter, $h=0.742 \pm 0.036$, based on the results of \citet{Riess09}.\footnote{Strictly speaking, this result should be interpreted as a constraint on the luminosity distance to $z \sim 0.04$, as noted by \citet{Reid09a}. However, as those authors conclude, the distinction between that approach and a straightforward prior on $h$ is very small in practice.}

\section{Analysis Method and Systematics} \label{sec:method}

Our results are obtained via Markov Chain Monte Carlo (MCMC), employing the Metropolis sampler embedded in the {\sc cosmomc} code of \citet{Lewis02}.\footnote{\url{http://cosmologist.info/cosmomc/}} The May 2008 {\sc cosmomc} release includes the 5-year \WMAP{} and Union supernova data and analysis codes; an additional module implementing the \fgas{} analysis has also been publicly released (\citealt*{Rapetti05}; \citealt{Allen08}).\footnote{\url{http://www.stanford.edu/~drapetti/fgas_module/}} Further modifications were made to include the likelihood codes for the XLF and BAO data. The CMB and matter power spectrum calculations were performed using the {\sc camb} package of \citet*{Lewis00}.\footnote{\url{http://www.camb.info}}

When analyzing the CMB data, we marginalize over a plausible range in the amplitude of the Sunyaev-Zel'dovich signal due to galaxy clusters ($0<\mysub{A}{SZ}<2$; introduced by \citealt{Spergel07}). Our analysis of the Union supernova sample of \citet{Kowalski08} includes their treatment of systematic uncertainties, which accounts for the effects of Malmquist bias and uncertainties in lightcurve fitting and photometry (among others).

The method used to analyze cluster \fgas{} data is described in full by \citet{Allen08}. It incorporates generous systematic allowances for instrument calibration (10 per cent), non-thermal pressure support \citep[10 per cent,][]{Nagai07}, the depletion of baryons in clusters with respect to the cosmic mean (20 per cent), and evolution with redshift of the baryonic and stellar content of clusters (10 and 20 per cent).

The analysis of the XLF data is detailed in \cosmopaper{}. The method combines cluster survey data with follow-up observations in an internally consistent way, rigorously accounting for the effects of Malmquist and Eddington biases and parameter degeneracies. Conservative systematic allowances are included to account for uncertainty in the predicted cluster mass function, the overall cluster survey completeness and purity, and instrument calibration.

As discussed in \secref{sec:background} and illustrated in \secrefs{sec:mnuresults} and \ref{sec:neffresults}, the XLF data contribute to constraints on \Mnu{} and \Neff{} through their measurement of $\sigma_8$. The posterior uncertainty on this measurement (from the XLF alone) is $\sim 6$ per cent, and is dominated by systematic uncertainty, as described in \cosmopaper{}. Specifically, the uncertainty in $\sigma_8$ is determined by the accuracy with which cluster masses can be measured. As detailed in \scalingpaper{}, we do not directly infer cluster masses at $r_{500}$ by assuming hydrostatic equilibrium of the intracluster medium, a procedure which, when applied to a typical cluster, introduces a large and highly variable bias \citep{Faltenbacher05,Rasia06,Nagai07}. Instead, we use the gas mass at $r_{500}$, which can be measured without significant bias, as a proxy for total mass. The proxy relation is calibrated using the \fgas{} data of \citet{Allen08}, which effectively consist of gas mass and total mass measurements for the subset of clusters where the hydrostatic assumption is applicable (the most massive, dynamically relaxed clusters). Of the systematic allowances described above, the most relevant for this procedure are the allowances for non-thermal support and instrument calibration in the \fgas{} analysis. In addition, we account for uncertainty in the difference in \fgas{} between $r_{2500}$ (as measured by Allen et~al.) and $r_{500}$, as well as possible scatter in \fgas{} from cluster to cluster (see \scalingpaper{}). Ultimately, both our estimates of individual masses and the mean cluster mass scale include a systematic error budget of $\sim 15$ per cent.

\section{Limits on \Mnu{}} \label{sec:mnuresults}

\subsection{Simple models} \label{sec:mnusimple}

We first consider the case of a spatially flat \LCDM{} cosmology with non-zero neutrino mass. For this model, the joint 68.3 and 95.4 per cent confidence regions in the \Mnu{}-$\sigma_8$ plane from the combination of CMB, \fgas{}, SNIa and BAO data appear in the left panel of \figref{fig:mslcdm} (blue contours). The 95.4 per cent confidence upper bound on \Mnu{} from these data (including the systematic allowances described in \secref{sec:method}) is $\Mnu<0.61\eV$ (\tabref{tab:msdegen}), a marginal improvement over that obtained by \citet{Komatsu09} using only \WMAP{}, SNIa and BAO data.

However, it is clear in the figure that a tight constraint on $\sigma_8$ can improve the limits on \Mnu{}. The upper limit obtained by including the XLF data, which provide such a constraint, is $\Mnu<0.33\eV$ (corresponding to the gold contours in \figref{fig:mslcdm}), nearly a factor of two improvement.

We note that when the results excluding the XLF data are combined with the Gaussian prior $\sigma_8=0.82 \pm 0.05$, based on the XLF results for a flat \LCDM{} cosmology with $\ns=0.95$ (\cosmopaper{}), the limits on \Mnu{} obtained are virtually identical to those from the full combination of data, $\Mnu<0.33\eV$. From this we conclude that the XLF data currently contribute to the bound on \Mnu{} only through their ability to constrain $\sigma_8$; the effect of non-zero neutrino mass on the (time-dependent) growth of structure is not detected in the present data.

\subsection{Extended models} \label{sec:mnuexten}

\begin{figure*}
  \centering
  \includegraphics[scale=\figscl]{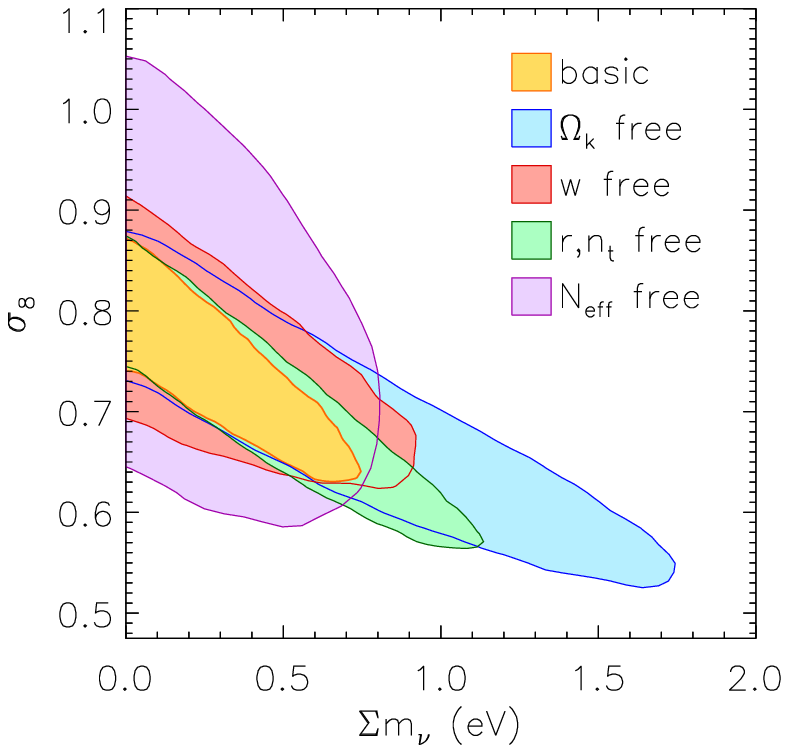}
  \hspace{0.5cm}
  \includegraphics[scale=\figscl]{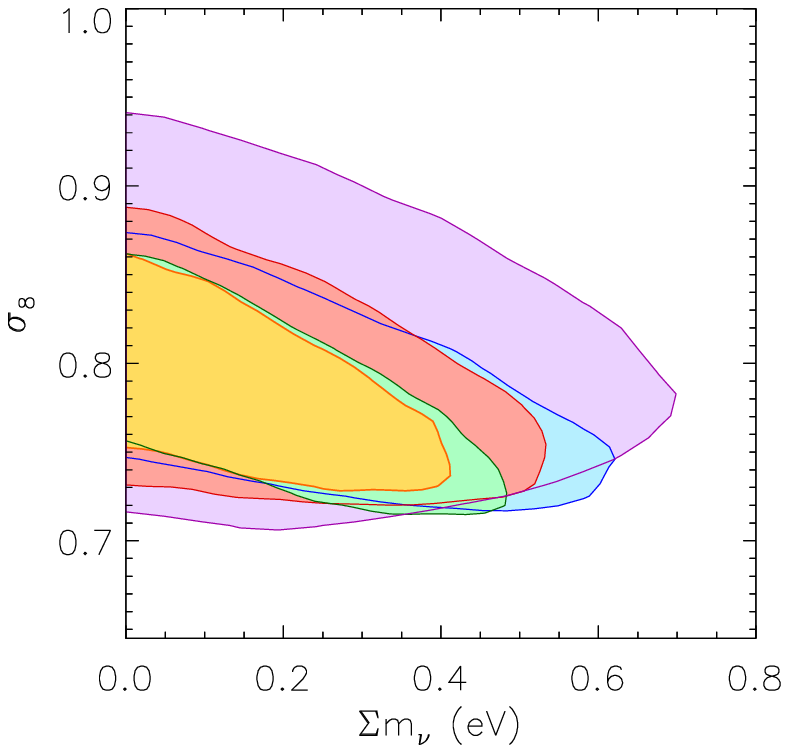}
  \caption{[{\it preprint note: a monochrome version of this figure appears after the references, as \figref{fig:msdegenbw}}] Joint 95.4 per cent confidence regions for \Mnu{} and $\sigma_8$ for various cosmological models. Yellow contours correspond to the basic \LCDM+\Mnu{} model, blue contours are marginalized over \Omegak{}, red contours over $w$, green over $r$ and \nt{}, and purple over \Neff{}. The left panel shows constraints obtained from the combination of CMB, \fgas{}, SNIa and BAO data; the right panel shows results that include the XLF in addition to those data. No external prior on $H_0$ is used.}
  \label{fig:msdegen}
\end{figure*}

\begin{table*}
  \centering
  \caption{Constraints on the species-summed neutrino mass and other model parameters, including conservative systematic allowances. Note that limits on \Mnu{} and $r$ are listed at 95.4 per cent confidence, while all others are 68.3 per cent confidence. Parameters with a single value listed were fixed at that value. Constraints on \Omegam{} do not appear in the table, as they were extremely similar for all models, consistent with $\Omegam=0.26 \pm 0.015$. The first column indicates whether the XLF data were included in the fit; the combination of CMB, \fgas{}, SNIa and BAO data is used in all cases. The results in this table do not use an external prior on $H_0$. The use of such a prior, in addition to the XLF data, greatly improves constraints on both \Neff{} and \Mnu{} when \Neff{} is a free parameter (see \tabref{tab:nhdegen}).}
  \begin{tabular}{cccccccl}
    \hline
    XLF & $\sigma_8$ & \Omegak{} & $w$ & $r$ & \nt{} & \Neff{} & \Mnu{} (eV) \\
    \hline
            & $0.77 \pm 0.05$ & $0$ & $-1$ & $0$ & $0$ & $3.046$ & $<0.61$ \\
    $\surd$ & $0.79 \pm 0.03$ & $0$ & $-1$ & $0$ & $0$ & $3.046$ & $<0.33$ \vspace{2.0ex} \\
            & $0.72 \pm 0.08$ & $0.009 \pm 0.008$ & $-1$ & $0$ & $0$ & $3.046$ & $<1.44$ \\
    $\surd$ & $0.79 \pm 0.03$ & $0.004 \pm 0.006$ & $-1$ & $0$ & $0$ & $3.046$ & $<0.50$ \vspace{2.0ex} \\
            & $0.76 \pm 0.05$ & $0$ & $-1.02 \pm 0.08$ & $0$ & $0$ & $3.046$ & $<0.72$ \\
    $\surd$ & $0.79 \pm 0.03$ & $0$ & $-1.03 \pm 0.07$ & $0$ & $0$ & $3.046$ & $<0.43$ \vspace{2.0ex} \\
            & $0.76 \pm 0.06$ & $0$ & $-1$ & $<0.33$ & $0.3^{+0.8}_{-0.7}$ & $3.046$ & $<0.92$ \vspace{0.75ex} \\
    $\surd$ & $0.79 \pm 0.03$ & $0$ & $-1$ & $<0.22$ & $0.5^{+0.7}_{-0.8}$ & $3.046$ & $<0.38$ \vspace{2.0ex} \\
            & $0.80 \pm 0.09$ & $0$ & $-1$ & $0$ & $0$ & $3.4_{-1.3}^{+2.0}$ & $<0.67$ \vspace{0.75ex} \\
    $\surd$ & $0.80 \pm 0.04$ & $0$ & $-1$ & $0$ & $0$ & $3.6_{-1.0}^{+1.4}$ & $<0.54$ \vspace{2.0ex} \\
            & $0.69 \pm 0.08$ & $0.008 \pm 0.009$ & $-1$ & $<0.29$ & $0.5^{+0.8}_{-0.8}$ & $3.046$ & $<1.60$ \vspace{0.75ex} \\
    $\surd$ & $0.79 \pm 0.03$ & $0.003 \pm 0.006$ & $-1$ & $<0.19$ & $0.1^{+0.8}_{-0.8}$ & $3.046$ & $<0.49$ \vspace{2.0ex} \\
            & $0.72 \pm 0.10$ & $0.010 \pm 0.009$ & $-1$ & $0$ & $0$ & $3.4_{-1.2}^{+2.0}$ & $<1.84$ \vspace{0.75ex} \\
    $\surd$ & $0.79 \pm 0.04$ & $0.007 \pm 0.007$ & $-1$ & $0$ & $0$ & $3.6_{-0.9}^{+1.7}$ & $<1.14$ \vspace{2.0ex} \\
            & $0.79 \pm 0.10$ & $0$ & $-1$ & $<0.43$ & $0.2^{+0.6}_{-0.7}$ & $4.5_{-2.0}^{+2.2}$ & $<1.46$ \vspace{0.75ex} \\
    $\surd$ & $0.81 \pm 0.05$ & $0$ & $-1$ & $<0.33$ & $0.2^{+0.7}_{-0.8}$ & $4.5_{-1.6}^{+1.4}$ & $<0.95$ \vspace{2.0ex} \\
            & $0.71 \pm 0.10$ & $0.008 \pm 0.009$ & $-1$ & $<0.36$ & $0.4^{+0.6}_{-0.7}$ & $4.5_{-2.1}^{+2.2}$ & $<2.34$ \vspace{0.75ex} \\
    $\surd$ & $0.80 \pm 0.04$ & $0.004 \pm 0.008$ & $-1$ & $<0.30$ & $0.1^{+0.8}_{-0.7}$ & $4.0_{-1.1}^{+1.9}$ & $<1.31$ \\
    \hline
  \end{tabular}
  \label{tab:msdegen}
\end{table*}

Next, we consider how the neutrino mass limits are affected when additional degrees of freedom are introduced in the cosmological model. Several possibilities are worth considering here. To begin with, the model for dark energy may be generalized; the simplest ways to accomplish this by introducing a single additional free parameter are through the spatial curvature (i.e. by not linking the dark energy density directly to the matter density) and through the equation of state, retaining the assumption of spatial flatness in the latter case. The flat \LCDM{} case corresponds to effective curvature density $\Omegak=0$ and equation of state $w=-1$. Since current data do not support departures from the flat \LCDM{} model either through $\Omegak \neq 0$ or $w \neq -1$, we consider these cases separately rather than introducing both parameters simultaneously. Secondly, we consider the effect of primordial tensor modes, parametrized through the tensor-to-scalar ratio, $r$. We additionally marginalize over the tensor spectral index, \nt{}, rather than assuming a particular model of inflation where \nt{} would be linked to $r$ via a consistency relation. Finally, we consider the effect of varying \Neff{} on the bounds obtained for \Mnu{}, leaving a discussion of the constraints on \Neff{} itself for \secref{sec:neffresults}.

The effects of adding each of these degrees of freedom individually to the \LCDM{}+\Mnu{} model are displayed in \figref{fig:msdegen} (95.4 per cent confidence regions only). The left panel shows results from the combination of CMB, SNIa, BAO and \fgas{} data. Here it is clear that degeneracies exist between \Mnu{} and some of the additional parameters. In particular, the presence of primordial tensor modes ($\Mnu<0.92\eV$) or curvature ($\Mnu<1.44\eV$) significantly degrades the limits on neutrino mass relative to the simpler model ($\Mnu<0.61\eV$) by lengthening the major axis of the confidence region. In constrast, while \Neff{} has a small effect on the \Mnu{} limit ($\Mnu<0.67\eV$), it significantly degrades the constraints on other cosmological parameters, including $\sigma_8$, resulting in the much thicker confidence region in the figure (see also \secref{sec:neffsimple}). The equation of state of dark energy is an intermediate case which increases the bound on the neutrino mass somewhat ($\Mnu<0.75\eV$), while also degrading the constraints on other parameters slightly. Marginalized confidence intervals on parameters of interest for these cases can be found in \tabref{tab:msdegen}.

The right panel of \figref{fig:msdegen} shows the equivalent tests when the XLF data are included in the analysis. In this case, the weakest constraint occurs when \Neff{} is free ($\Mnu<0.54\eV$); this is sensible, since it had the smallest correlation with $\sigma_8$ a priori (left panel). In the other cases shown, limits on \Mnu{} are improved by a factor of 2--3. In the most general model considered, where curvature, tensors and additional relativistic species are all marginalized over, the full combination of data yields $\Mnu<1.31\eV$, compared with $\Mnu<2.34\eV$ when the XLF is not used (\tabref{tab:msdegen}).
\vfill

\section{Constraints on \Neff{}} \label{sec:neffresults}

\subsection{Simple models} \label{sec:neffsimple}

\begin{figure}
  \centering
  \includegraphics[scale=\figscl]{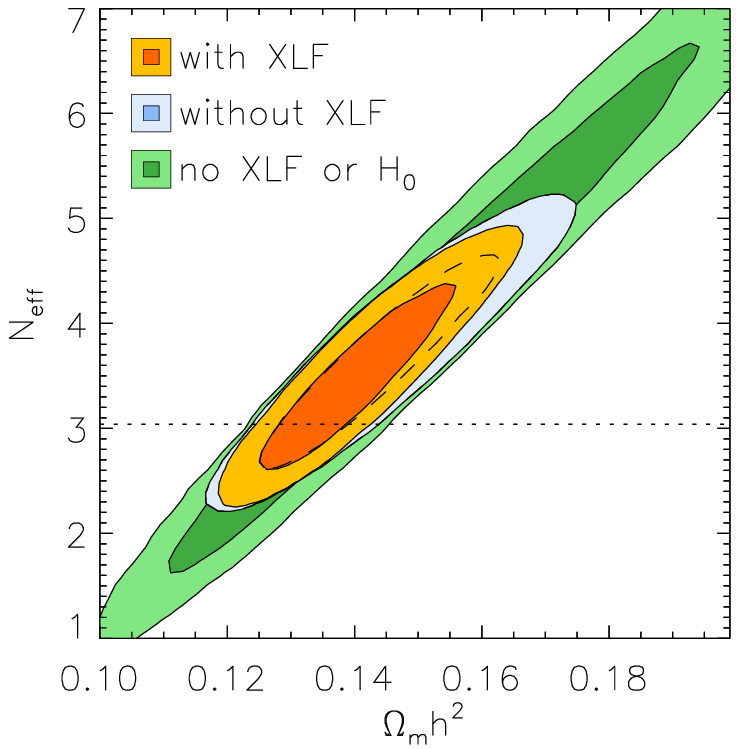}
  \caption{Joint 68.3 and 95.4 per cent confidence regions on $\Omegam h^2$ and \Neff{} from the combination of CMB, \fgas{}, SNIa and BAO data with a direct measurement of $H_0$ (blue), and the same data with the addition of the XLF (gold). Also shown are the constraints from the former combination of data, but without the prior on $H_0$ (green); this illustrates the strong sensitivity of the \Neff{} results to the Hubble parameter. The dotted, horizontal line indicates the standard value of \Neff{}, 3.046.}
  \label{fig:hnlcdm}
\end{figure}

Joint constraints on \Neff{} and $\Omegam h^2$ are shown in \figref{fig:hnlcdm} for a flat \LCDM{} cosmology. The strong correlation in this plane is due to the fact that the  CMB data constrain directly \zeq{}, which is a degenerate combination of \Neff{} and $\Omegam h^2$, as discussed previously in \secref{sec:background}. The combination of data that we use places tight constraints on \Omegam{}, so significant improvement can be obtained by incorporating a direct measurement of the Hubble constant; throughout this section we use a Gaussian prior, $h=0.742 \pm 0.036$, based on the results of \citet{Riess09}. The constraints thus obtained are listed in \tabref{tab:nhdegen}.

\begin{figure}
  \centering
  \includegraphics[scale=\figscl]{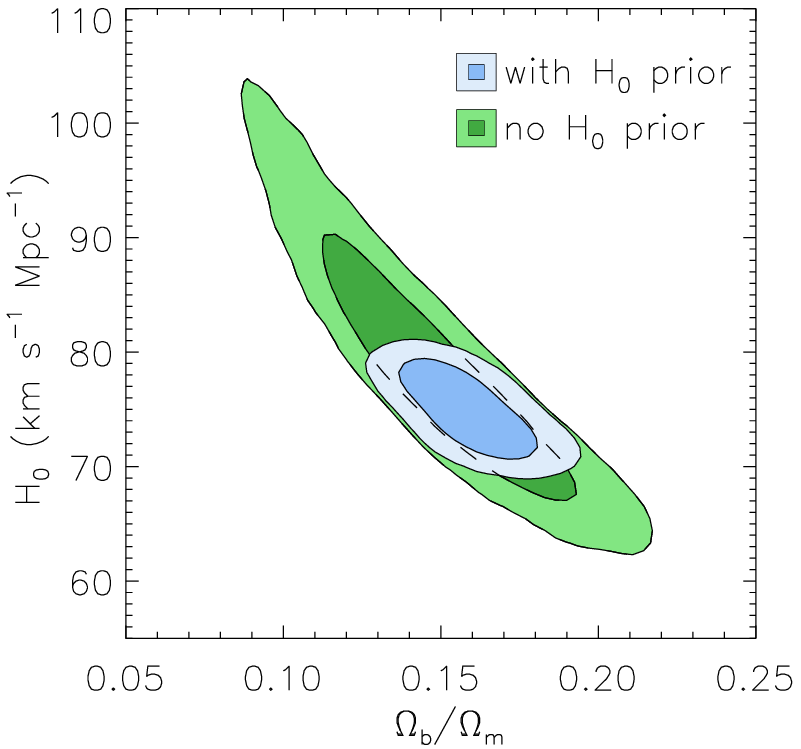}
  \caption{Joint constraints on $\Omegab/\Omegam$ and $H_0$ when \Neff{} is free. Green contours show results obtained from the combination of CMB, \fgas{}, SNIa and BAO data; despite the inclusion of SNIa and BAO data, the strong degeneracy in this plane that occurs in both \fgas{} and CMB analyses (when \Neff{} is free) remains evident. The addition of a measurement of the Hubble parameter at the 5 per cent level (blue contours) significantly improves the results.}
  \label{fig:barfrac}
\end{figure}

We note that, ordinarily, the combination of \fgas{} and CMB data provides a tight constraint on $H_0$ \citep{Allen08}; however, this is not the case when \Neff{} is free. The reason why can be seen in \figref{fig:barfrac}. The green contours in the figure show constraints obtained from combining the CMB, \fgas{}, SNIa and BAO data (without a prior on $H_0$). A correlation in this plane occurs naturally in the \fgas{} analysis, since X-ray observations of clusters measure a degenerate combination of the cosmic baryon fraction, $\Omegab/\Omegam$, and distance. When \Neff{} is fixed, CMB data place tight constraints on the baryon fraction, though not on the Hubble parameter, and so the combination of CMB and \fgas{} data produces tight constraints on both $H_0$ and $\Omegab/\Omegam$. When \Neff{} is free, however, the CMB constraints in this plane are degenerate along nearly the same axis as the \fgas{} constraints,\footnote{Respectively, the quantities $\Omegam h^2$ and $\Omegam h^{1.5}$ are degenerate in the analysis of CMB and \fgas{} data.} and so the inclusion of additional, independent distance measurements is necessary to place a constraint on $H_0$. The combination with a direct measurement of $H_0$ (blue contours) significantly improves matters.

\begin{figure}
  \centering
  \includegraphics[scale=\figscl]{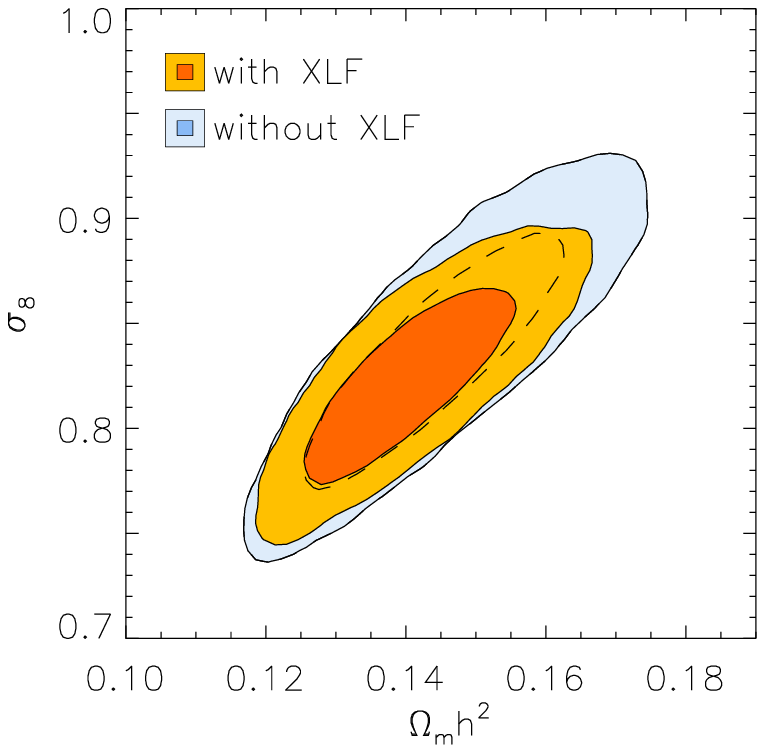}
  \caption{Joint 68.3 and 95.4 per cent confidence regions on $\Omegam h^2$ and $\sigma_8$ from the combination of CMB, \fgas{}, SNIa and BAO data with a direct measurement of $H_0$ (blue), and the same data with the addition of the XLF (gold) when \Neff{} is free. The tight constraint on $\sigma_8$ provided by the XLF data improves the determination of $\Omegam h^2$, which translates to an improved constraint on \Neff{} (\figref{fig:hnlcdm}).}
  \label{fig:hslcdm}
\end{figure}

As \figref{fig:hnlcdm} shows, the addition of the XLF data improves the constraint on \Neff{} somewhat, from $\Neff=3.6_{-0.6}^{+0.7}$ to $\Neff=3.4_{-0.5}^{+0.6}$ (68.3 per cent confidence). The mechanism for this improvement is a degeneracy between $\Omegam h^2$ and $\sigma_8$, shown in \figref{fig:hslcdm}, which the XLF data reduce through their constraint on $\sigma_8$.

\subsection{Extended models}

\begin{table*}
  \centering
  \caption{Constraints on the neutrino effective number and other model parameters. Note that limits on \Mnu{} and $r$ are listed at 95.4 per cent confidence, while all others are 68.3 per cent confidence. Parameters with a single value listed were fixed at that value. The first two columns respectively indicate whether the prior on $H_0$ and the XLF data were included in the fit; the combination of CMB, \fgas{}, SNIa and BAO data is used in all cases.}
  \begin{tabular}{cccccccc}
    \hline
    $H_0$ & XLF & $\sigma_8$ & \Omegak{} & $r$ & \nt{} & \Neff{} & \Mnu{} (eV)\\
    \hline
            &         & $0.86 \pm 0.07$ & $0$ & $0$ & $0$ & $3.8_{-1.5}^{+1.9}$ & $0$ \vspace{0.75ex} \\
    $\surd$ &         & $0.83 \pm 0.04$ & $0$ & $0$ & $0$ & $3.6_{-0.6}^{+0.7}$ & $0$ \vspace{0.75ex} \\
    $\surd$ & $\surd$ & $0.82 \pm 0.03$ & $0$ & $0$ & $0$ & $3.4_{-0.5}^{+0.6}$ & $0$ \vspace{2.0ex} \\
            &         & $0.85 \pm 0.07$ & $0.002 \pm 0.006$ & $0$ & $0$ & $3.4_{-1.3}^{+1.8}$ & $0$ \vspace{0.75ex} \\
    $\surd$ &         & $0.83 \pm 0.04$ & $0.001 \pm 0.005$ & $0$ & $0$ & $3.4_{-0.7}^{+0.8}$ & $0$ \vspace{2.0ex} \\
            &         & $0.88 \pm 0.08$ & $0$ & $<0.25$ & $0.1^{+0.9}_{-0.7}$ & $4.2_{-1.7}^{+2.6}$ & $0$ \vspace{0.75ex} \\
    $\surd$ &         & $0.83 \pm 0.04$ & $0$ & $<0.20$ & $0.3^{+0.9}_{-0.9}$ & $3.6_{-0.8}^{+0.7}$ & $0$ \vspace{2.0ex} \\
            &         & $0.80 \pm 0.09$ & $0$ & $0$ & $0$ & $3.4_{-1.3}^{+2.0}$ & $<0.67$ \vspace{0.75ex} \\
    $\surd$ &         & $0.79 \pm 0.05$ & $0$ & $0$ & $0$ & $3.5_{-0.4}^{+0.9}$ & $<0.64$ \vspace{0.75ex} \\
    $\surd$ & $\surd$ & $0.80 \pm 0.03$ & $0$ & $0$ & $0$ & $3.5_{-0.5}^{+0.7}$ & $<0.48$ \vspace{2.0ex} \\
            &         & $0.72 \pm 0.10$ & $0.010 \pm 0.009$ & $0$ & $0$ & $3.4_{-1.2}^{+2.0}$ & $<1.84$ \vspace{0.75ex} \\
    $\surd$ &         & $0.73 \pm 0.08$ & $0.008 \pm 0.009$ & $0$ & $0$ & $3.4_{-0.6}^{+0.9}$ & $<1.58$ \vspace{0.75ex} \\
    $\surd$ & $\surd$ & $0.79 \pm 0.03$ & $0.005 \pm 0.006$ & $0$ & $0$ & $3.5_{-0.5}^{+0.8}$ & $<0.72$ \vspace{2.0ex} \\
            &         & $0.79 \pm 0.10$ & $0$ & $<0.43$ & $0.2^{+0.6}_{-0.7}$ & $4.5_{-2.0}^{+2.2}$ & $<1.46$ \vspace{0.75ex} \\
    $\surd$ &         & $0.76 \pm 0.07$ & $0$ & $<0.34$ & $0.2^{+0.7}_{-0.7}$ & $3.7_{-0.9}^{+0.7}$ & $<1.14$ \vspace{0.75ex} \\
    $\surd$ & $\surd$ & $0.80 \pm 0.04$ & $0$ & $<0.23$ & $0.3^{+0.9}_{-0.8}$ & $3.6_{-0.6}^{+0.8}$ & $<0.55$ \vspace{2.0ex} \\
            &         & $0.71 \pm 0.10$ & $0.008 \pm 0.009$ & $<0.36$ & $0.4^{+0.6}_{-0.7}$ & $4.5_{-2.1}^{+2.2}$ & $<2.34$ \vspace{0.75ex} \\
    $\surd$ &         & $0.71 \pm 0.09$ & $0.006 \pm 0.009$ & $<0.33$ & $0.4^{+0.7}_{-0.7}$ & $3.7_{-0.8}^{+0.7}$ & $<1.75$ \vspace{0.75ex} \\
    $\surd$ & $\surd$ & $0.80 \pm 0.04$ & $0.002 \pm 0.007$ & $<0.29$ & $0.2^{+0.8}_{-0.7}$ & $3.7_{-0.7}^{+0.7}$ & $<0.70$ \\
    \hline
  \end{tabular}
  \label{tab:nhdegen}
\end{table*}

The addition of nuisance parameters in the form of curvature, tensors, and non-zero neutrino mass tends to weaken the constraints on \Neff{} and $H_0$ along their primary degeneracy axis, similarly to what was observed in \secref{sec:mnuexten}. \tabref{tab:nhdegen} lists the constraints obtained with and without the $H_0$ prior when various nuisance parameters are marginalized over, as well as when the XLF data are included in the fit. As the table shows, the inclusion of the prior on $H_0$ effectively eliminates the degeneracies between \Neff{} and the nuisance parameters; in particular, the constraint on \Neff{} obtained when marginalizing over curvature, tensors and neutrino mass simultaneously, $\Neff=3.7_{-0.8}^{+0.7}$, is very similar to those obtained when the nuisance parameters are fixed, $\Neff=3.6_{-0.6}^{+0.7}$. With so much extra freedom in the model, the inclusion of the XLF data produces only a small improvement in \Neff{} in that case.

\section{Prospects for improvement} \label{sec:improve}

An interesting question is where future improvements in the determination of \Mnu{} and \Neff{} from cosmological data will come from. One simple way of addressing this is to see what other cosmological parameters are most correlated with the parameters of interest in the current results. We consider the simple cases of a \LCDM{} model where either \Mnu{} or \Neff{} is additionally free, as well as the more general case with both parameters free in addition to curvature and tensor modes.

\subsection{Improvements on \Mnu{}}

When only \Mnu{} is additionally free, it remains most degenerate with $\sigma_8$, despite the tight constraints placed on this parameter by current data, with a correlation coefficient of $\rho=-0.63$. In the more general model (with \Mnu{}, \Neff{}, \Omegak{}, $r$ and \nt{} free), the degeneracy with $\sigma_8$ is also relatively important ($\rho=-0.51$), but degeneracies with $\Omegam h^2$ ($\rho=0.60$) and \Omegak{} ($\rho=0.48$) appear at approximately the same level.

The situation will change, however, with the availability of improved CMB data from {\it Planck}, and ultimately from future high-resolution CMB polarization and lensing observations. These new data will tighten constraints on a number of parameters that contribute to the width of the degenerate region in the \Mnu{}-$\sigma_8$ plane,\footnote{In addition to the curvature and tensor parameters discussed in this paper, standard cosmological parameters such as $\Omegam h^2$, $\tau$, \mysub{n}{s} and the high-redshift power spectrum amplitude all contribute to the width.} (\figrefs{fig:mslcdm} and \ref{fig:msdegen}; \citealt*{Kaplinghat03,Bashinsky04,Colombo09,dePutter09}) resulting in a stronger correlation of \Mnu{} with $\sigma_8$.

Using a simulated {\it Planck} data set produced using the {\sc FuturCMB} code of \citet{Perotto06}, we project that such data, combined with a 2~per cent constraint on $\sigma_8$, will produce a limit $\Mnu<0.18\eV$ for minimal neutrino mass in the normal hierarchy ($\Mnu=0.056\eV$). Alternatively, $\Mnu=0$ would be ruled out at approximately $1\sigma$ significance for minimal neutrino mass in the inverted hierarchy ($\Mnu=0.095\eV$). A constraint on $\sigma_8$ at this level may be possible in the near term from galaxy cluster observations, primarily by better understanding cluster mass measurements (e.g. \citealt{Mahdavi08,Zhang09}; von~der~Linden et~al., in preparation), as well as from improved measurements of the CMB power spectrum at multipoles $\sim 2000$ \citep{Lueker09,Fowler10,Komatsu10}.

As growth of structure data become available at higher redshifts, the time-dependent effects of neutrino mass on structure formation will also provide a powerful and more direct constraint on \Mnu{}. Future cluster surveys may provide limits an order of magnitude better than current results \citep{Wang05}. Eventually, other cosmological observables, e.g. the cross-correlation of Lyman-alpha absorption with CMB convergence \citep{Vallinotto09}, may extend measurements of structure formation to even higher redshifts ($z \sim 2$).

\subsection{Improvements on \Neff{}}

Unsurprisingly, the correlation of \Neff{} with $\Omegam h^2$ is dominant in both the simple (\LCDM+\Neff{}) and more general (\Mnu{}, \Neff{}, \Omegak{}, $r$ and \nt{} free) models, respectively with $\rho=0.94$ and 0.87, although we note that in the simple case there is also a strong degeneracy with $\sigma_8$ ($\rho=0.73$). Thus, improved measurements of the Hubble parameter and/or mean matter density will be very significant for cosmological constraints on \Neff{}.

In particular, we consider the possibility of a direct constraint on $H_0$ at the 2 per cent level. Such a constraint could come from several avenues, including {\it Spitzer} (and ultimately the {\it James Webb Space Telescope}) observations of Cepheids and SNIa \citep{Freedman09,Riess09}, gravitational lensing time delays \citep[e.g.][]{Saha06,Orguri07,Coe09,Dobke09}, or the comparison of the Compton-$y$ parameter inferred from Sunyaev-Zel'dovich and X-ray observations of clusters (\citealt*{Schmidt04}; \citealt{Bonamente06}; \citealt*{Rapetti08}). The latter two naturally complement efforts to extend measurements of the growth of structure using clusters to larger samples and higher redshifts. Even with no other improvements in cosmological data, such a measurement (we take $h=0.72 \pm 0.0144$, for concreteness) would provide a constraint $\Neff=3.04 \pm 0.35$ in the simple \LCDM+\Neff{} model, or $\Neff=3.14 \pm 0.48$ in the general model (parameters listed above; 68.3 per cent confidence). Since \Neff{} and \Mnu{} are degenerate in this model, the upper limit on \Mnu{} would also be improved, from 0.70 to 0.55\eV.

\section{Conclusion} \label{sec:conclusion}

We have applied measurements of the growth of massive galaxy clusters (detailed in Papers~I and II), in combination with other cosmological data, to the problem of constraining the species-summed neutrino mass, \Mnu{}, and effective number, \Neff{}.

Our results show that a robust measurement of $\sigma_8$ from clusters significantly improves limits on \Mnu{} from current data, and reduces their sensitivity to assumptions about the cosmological model. In a simple \LCDM+\Mnu{} cosmology, the addition of the XLF data improves the 95.4 per cent confidence upper limit on \Mnu{} to 0.33\eV{}, compared with 0.61\eV{} from the combination of CMB, \fgas, SNIa and BAO data. In a more general model, marginalized over spatial curvature, primordial tensors and the effective number of neutrinos, and incorporating a prior on the Hubble parameter, the XLF data improve the limit from 1.75\eV{} to 0.7\eV{}. The results indicate that this improvement is due entirely to the ability of the cluster data to constrain $\sigma_8$ (at $z=0$); while measurements of the ($z$-dependent) growth of structure in principle contain additional information about \Mnu{}, current data are not sufficient to exploit it.

The cluster data additionally improve constraints on the effective number of relativistic species, from $\Neff=3.6_{-0.6}^{+0.7}$ to $\Neff=3.4_{-0.5}^{+0.6}$ in a simple \LCDM+\Neff{} model (68.3 per cent confidence). For the more general model, including curvature, tensors and neutrino mass, only a marginal improvement is observed.

The results obtained here are compatible with other recent estimates based on galaxy clusters \citep{Reid09a,Vikhlinin09a}, reflecting the good agreement in recent $\sigma_8$ constraints based on X-ray and optically selected clusters \citep[\cosmopaper{};][]{Henry09,Rozo10}. Although current cosmological data are not sufficient to rule out the existence of sterile neutrino species or distinguish between the normal and inverted mass hierarchies, they continue to provide some of the tightest constraints available, in particular on the mass scale.

We also consider the prospects for further improvement, finding that a few per cent level measurement of the Hubble parameter will significantly improve the constraints on models where \Neff{} is free. A similarly improved determination of $\sigma_8$, combined with improved CMB data from {\it Planck}, will further tighten limits on the neutrino mass scale,\footnote{We note that the 7-year \WMAP{} results, which were released while this work was in revision, already offer some improvement. Specifically, \citet{Komatsu10} find $\Mnu<0.58\eV$ by combining 7-year \WMAP{} data with the \citet{Riess09} $H_0$ prior and the BAO results of \citet{Percival10}, a noticeable improvement over the limit $\Mnu<0.66\eV$ obtained by \citet{Sekiguchi09} from the same auxiliary data sets combined with WMAP5.} perhaps providing the first hints of non-zero mass from cosmological data. More precise measurements of the growth of structure at late times, and the extension of such data to higher redshifts, will provide a powerful, new mechanism to constrain neutrino mass.

\section*{Acknowledgments} \label{sec:acknowledgements}

We are grateful to Harald Ebeling and Alex Drlica-Wagner for their contributions to this series of papers, and to Glenn Morris, Stuart Marshall and the SLAC unix support team for technical support. We also thank Giorgio Gratta and Naoko Kurahashi for useful discussions. Calculations were carried out using the KIPAC XOC and Orange compute clusters at the SLAC National Accelerator Laboratory and the SLAC Unix compute farm. We acknowledge support from the National Aeronautics and Space Administration (NASA) through Chandra Award Numbers DD5-6031X, GO7-8125X and GO8-9118X, issued by the Chandra X-ray Observatory Center, which is operated by the Smithsonian Astrophysical Observatory for and on behalf of NASA under contract NAS8-03060. This work was supported in part by the U.S. Department of Energy under contract number DE-AC02-76SF00515. AM was supported by a Stanford Graduate Fellowship and an appointment to the NASA Postdoctoral Program, administered by Oak Ridge Associated Universities through a contract with NASA.

\begin{figure*}
  \centering
  \includegraphics[scale=\figscl]{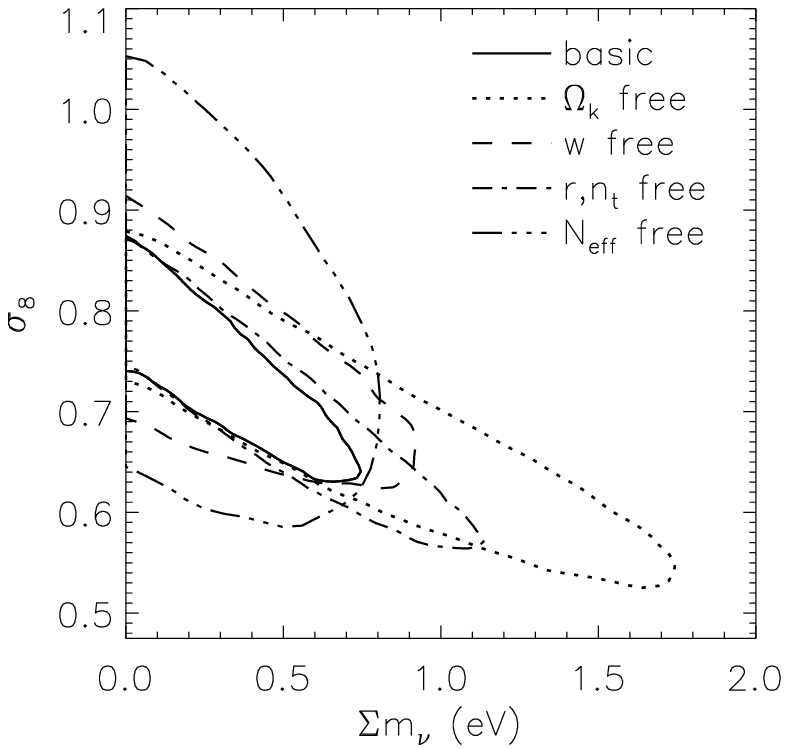}
  \hspace{0.5cm}
  \includegraphics[scale=\figscl]{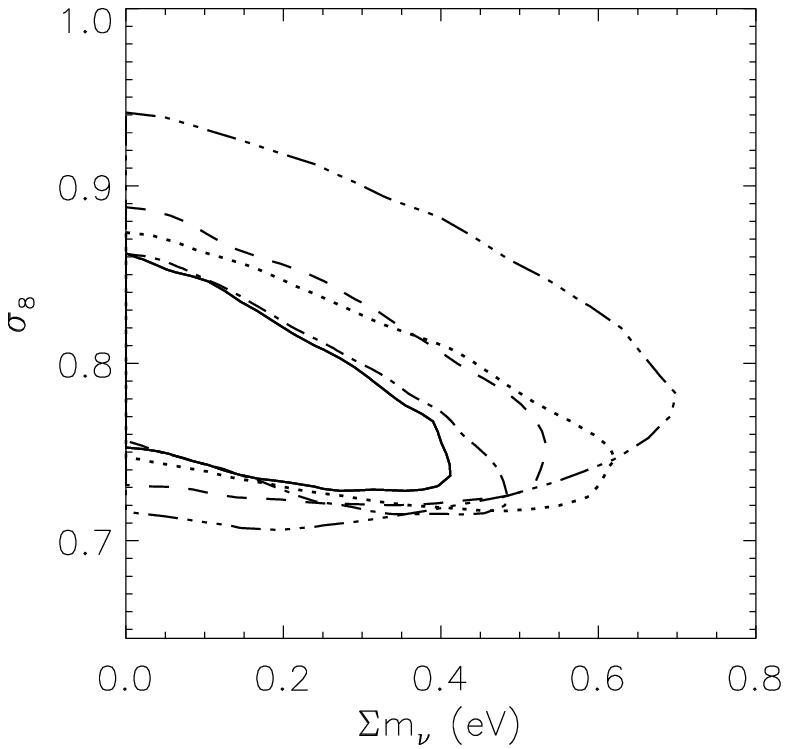}
  \caption{(Black and white version of \figref{fig:msdegen}.) Joint 95.4 per cent confidence regions for \Mnu{} and $\sigma_8$ for various cosmological models. Solid contours correspond to the basic \LCDM+\Mnu{} model, dotted contours are marginalized over \Omegak{}, dashed contours over $w$, dot-dashed over $r$ and \nt{}, and 3-dot-dashed over \Neff{}. The left panel shows constraints obtained from the combination of CMB, \fgas{}, SNIa and BAO data; the right panel shows results that include the XLF in addition to those data. No external prior on $H_0$ is used.}
  \label{fig:msdegenbw}
\end{figure*}

\bibliographystyle{mnras}
\def \aap {A\&A} 
\def \statisci {Statis. Sci.}
\def \physrep {Phys. Rep.}
\def \pre {Phys.\ Rev.\ E}
\def \sjos {Scand. J. Statis.} 
\def \jrssb {J. Roy. Statist. Soc. B} 
\def \pan {Phys. Atom. Nucl.} 
\def \epja {Eur. Phys. J. A} 
\def \epjc {Eur. Phys. J. C} 
\def \jcap {J. Cosmology Astropart. Phys.} 
\def \ijmpd {Int.\ J.\ Mod.\ Phys.\ D}
\def \araa {ARA\&A}
\def \aj {AJ}
\def \apj {ApJ}
\def \apjl {ApJL}
\def \apjs {ApJS}
\def \mnras {MNRAS}
\def \nat {Nat}
\def \pasj {PASJ}
\def \gca {Geochim.\ Cosmochim.\ Acta}
\def \npa {Nucl.\ Phys.\ A}
\def \plb {Phys.\ Lett.\ B}
\def \prc {Phys.\ Rev.\ C}
\def \prd {Phys.\ Rev.\ D}
\def \prl {Phys.\ Rev.\ Lett.}

\bsp
\label{lastpage}
\end{document}